\documentclass[aps,twocolumn,superscriptaddress,amsmath,showpacs,nofootinbib]{revtex4}

\usepackage{amsmath}
\usepackage{graphicx}
\usepackage{amssymb}
\usepackage{epsfig}
\usepackage{amsfonts}
\usepackage{color}

 \def\fig#1{{#1}}

\newcommand{\etal}{{\em et al.}}
\newcommand{\fra}[2]{{\textstyle\frac{#1}{#2}\,}}
\newcommand{\<}{\langle} \def\>{\rangle}
\newcommand{\psiT}{(|\psi\rangle\langle\psi|)^\Gamma} \def\Bop
{\mathcal{B'}_{\mathrm{CHSH}}}

\newcommand{\pla}{Phys. Lett. A~}
\newcommand{\job}{J. Opt. B: Quantum Semiclass. Opt.~}

\begin{document}

\title{Entanglement estimation from Bell inequality violation}

\author{Karol Bartkiewicz}
\email{bartkiewicz@jointlab.upol.cz} \affiliation{RCPTM,
Joint Laboratory of Optics of Palack\'y University and
Institute of Physics of Academy of Sciences of the Czech
Republic, 17. listopadu 12, 772 07 Olomouc, Czech Republic }
\affiliation{Faculty of Physics, Adam Mickiewicz University,
61-614 Pozna\'n, Poland}

\author{Bohdan Horst}

\affiliation{Faculty of Physics, Adam Mickiewicz University,
61-614 Pozna\'n, Poland}

\author{Karel Lemr}

\affiliation{RCPTM, Joint Laboratory of Optics of Palack\'y
University and Institute of Physics of Academy of Sciences of
the Czech Republic, 17. listopadu 12, 772 07 Olomouc, Czech
Republic }

\author{Adam Miranowicz}

\affiliation{Faculty of Physics, Adam Mickiewicz University,
61-614 Pozna\'n, Poland}

\begin{abstract}
It is well known that the violation of  Bell's inequality in
the form given by Clauser, Horne, Shimony, and Holt (CHSH) in
two-qubit systems requires entanglement, but not vice versa,
i.e., there are entangled states which do not violate the
CHSH inequality. Here we compare some standard entanglement
measures with violations of the CHSH inequality (as given by
the Horodecki measure) for two-qubit states generated by
Monte Carlo simulations. We describe states that have
extremal entanglement according to the negativity,
concurrence, and relative entropy of entanglement for a given
value of the CHSH violation. We explicitly find these
extremal states by applying the generalized method of
Lagrange multipliers based on the Karush-Kuhn-Tucker
conditions. The found minimal and maximal states define the
range of entanglement accessible for any two-qubit states
that violate the CHSH inequality by the same amount. We also
find extremal states for the concurrence versus negativity by
considering only such states which do not violate the CHSH
inequality. Furthermore, we describe an experimentally
efficient linear-optical method to determine the highest
Horodecki degree of the CHSH violation for arbitrary mixed
states of two polarization qubits. By assuming to have access
simultaneously to two copies of the states, our method
requires only six discrete measurement settings instead of
nine settings, which are usually considered.
\end{abstract}

\pacs{03.67.Mn, 03.65.Ud, 42.50.Dv}

\maketitle

\section{Introduction}

Since the seminal paper of Einstein, Podolsky and
Rosen~\cite{Einstein35}, there has been much interest in the two
seemingly interrelated phenomena of quantum entanglement and
nonlocality. Especially during the last three decades much
theoretical and experimental work has been done in order to
better understand the implications of these phenomena not
only in physics but even in biology and philosophy.

Quantum entanglement is nowadays relatively well
understood~\cite{Horodecki09}. It is defined as the
inseparability of quantum states and can be viewed as an
algebraic concept. Quantum nonlocality is more related to
experimental statistics. Namely, it can be considered as a
type of correlation between measurement outcomes, obtained
in spatially and temporally separated laboratories, that
cannot be explained by local hidden-variable theories.
Bell-type inequalities~\cite{Bell,Clauser69} are often used
to address this nonlocality quantitatively~\cite{Brunner13}.
In this paper we focus on the violation of Bell's
inequality in the form derived by Clauser, Horne, Shimony,
and Holt (referred to as the CHSH
inequality)~\cite{Clauser69}.

For two qubits, Bell inequalities can be violated
only if their states are entangled. However, as shown by
Werner~\cite{Werner89}, there are entangled states that can
still exhibit correlations which do not  violate
{any Bell inequality for any possible local measurements;
that is, unless a sequence of measurements, or several
copies, or other more sophisticated scenarios are
applied~\cite{Brunner13}. Note that Werner considered only
projective measurements, but his conclusions apply also to
the case of general measurements [positive operator-valued measures POVMs]~\cite{Brunner13}.}

Werner's states are defined as~\cite{Werner89}
\begin{eqnarray}
{\hat\rho }_{W}(p) &=& p|\Psi^{-}\rangle \langle\Psi^{-}|
+\fra{1-p}{4} I\otimes I, \label{eq:1}
\end{eqnarray}
which is a mixture of the singlet state
$|\Psi^{-}\rangle=(|01\rangle - |10\rangle)/\sqrt{2}$ and
the maximally mixed state $ I\otimes I$, where $I$ is the
single-qubit identity operator and the parameter $p\in [0,1]$. The Werner
states violate the CHSH inequality if and only if $1/\sqrt{2}<p\le 1$,
while they are entangled iff $1/3<p \le 1$. Thus, for $p\in
(1/3,1/\sqrt{2}],$ the Werner states are entangled and {they
satisfy} the CHSH inequality~\cite{Werner89}. These
properties of the Werner states can be easily revealed by
applying the Horodecki theorem~\cite{Horodecki95}.

Therefore, a natural question can be raised as to how much
entangled states can be without violating the CHSH inequality
or, more generally, for any fixed degree of the CHSH violation.
The intuitive guess is that different measures of
entanglement imply different answers for this question.

A degree of entanglement of two-qubit states can be described
by various entanglement measures
including~\cite{Horodecki09} (i) the relative entropy of
entanglement (REE)~\cite{Vedral97a}, which is a quantum
version of the Kullback-Leibler divergence; (ii) the
Peres-Horodecki negativity~\cite{Peres96}, which is a measure
of the entanglement cost under operations preserving the
positivity of partial transpose (PPT)~\cite{Audenaert03}; and
(iii) the Wootters concurrence~\cite{Wootters98}, a
measure of the entanglement of formation~\cite{Bennett1}. On
the other hand, the Horodecki theorem~\cite{Horodecki95}
enables not only testing the CHSH inequality violation but
also quantifying the degree of this violation for arbitrary
two-qubit states. This degree is often referred to as a
single-copy nonlocality measure~\cite{Brunner13}.

In this paper we shall use the listed measures of
entanglement to answer the question about the relation
between the CHSH violation and entanglement quantitatively.
In particular we find states that have extreme
entanglement for all the above-mentioned entanglement
measures for a given degree of the CHSH violation. For the
purpose of our optimization procedure we shall use the
so-called Karush-Kuhn-Tucker (KKT) conditions in a
generalized method of Lagrange multipliers, which provide
powerful tools for solving such optimization
problems~\cite{Boyd04}. We also use other tools for testing
the optimality of the states obtained, namely, the optimality
conditions for  the concurrence provided in
Ref.~\cite{Verstraete02} and Monte Carlo simulations.

Verstraete and Wolf~\cite{Verstraete02} found the
regions of possible extremal CHSH violation for a given
concurrence. This comparison is an important result, but it
does not indicate the regions of extremal CHSH violation for
other important entanglement measures including the
negativity and REE. For example, in contrast to the Verstraete-Wolf
results, pure states are \emph{not} extremal if the CHSH
violation is compared with the REE (for values \emph{not} too
close to 1) as recently shown in Ref.~\cite{Horst13}. Here
we give a deeper comparison of the REE and CHSH violation.
More importantly, we find the regions of the extremal
negativity for a given CHSH violation.

The inequivalence of such results for different
entanglement measures in comparison to the CHSH violation can
be understood by recalling that these measures have
fundamentally different physical meanings (as discussed in
Sec.~\ref{sec:def}) even for two-qubit states. Only in
special cases, including pure states,  the negativity and
concurrence become equal and equivalent to the REE. As an
example of such basic discrepancies, we will show explicitly
that these three entanglement measures do not necessarily
imply the same ordering of states even if the CHSH violation
is fixed at some value. Actually, as shown in
Ref.~\cite{Virmani00}, all ``good'' nonidentical asymptotic
entanglement measures (such as those studied in this paper) cannot
impose consistent orderings for all quantum states.  

We study the relation between the CHSH violation and
negativity (and other entanglement measures) for
\emph{arbitrary} two-qubit states analogously to the
comparisons of the CHSH violation with the
concurrence~\cite{Verstraete02,Derkacz05} and
REE~\cite{Horst13}. Note that many other comparative studies
of the concurrence and CHSH violation were limited to some
\emph{specific} classes of two-qubit states usually in a
dynamical context~\cite{Jakobczyk03,Miran04c,Derkacz04,
Liao07, Kofman08, Deng09, Mazzola10,Berrada11,Hu13}.

It is worth noting an increasing interest in developing
device-independent approaches to entanglement testing and
quantifying, which are based on various Bell inequality violations (see,
e.g., Ref.~\cite{Brunner13,Moroder13} and references
therein). For example, semi-device-independent upper and
lower bounds on the concurrence were studied in
Ref.~\cite{Liang11}, and device-independent lower bounds on
the negativity were found recently in Ref.~\cite{Moroder13}.
These approaches often correspond to testing only sufficient
conditions for the CHSH violation. They are,
however, beyond the scope of this work, which is focused on
the maximal violations of the CHSH inequality based on the
necessary and sufficient conditions as described by the
Horodecki measure. For example, Ref.~\cite{Moroder13}
employed some methods and results obtained in the studies of
matrices of moments for continuous variable systems related
to the criteria (i.e., witnesses instead of measures) of
entanglement~\cite{Shchukin05} and
nonclassicality~\cite{Vogel08}.

We focus on the CHSH inequality, although there are stronger
Bell inequalities as was shown for the Werner states by,
e.g., Vertesi~\cite{Vertesi08}. Note, however, that the CHSH
inequalities even though simple (there is a variety of other
Bell inequalities involving more measurement settings) are
very powerful since the stronger inequality requires at least
465 settings on each side for the Vertesi inequality. The
Werner states violate the Vertesi inequality for $p> 0.7056$,
while the CHSH inequality is violated, as already mentioned
for $p> 1/\sqrt{2} \sim 0.7071$ only. There are other Bell
inequalities that are not equivalent to the ones already mentioned. 
Notably, there is the \'Sliwa-Collins-Gisin
inequality~\cite{Sliwa03}, i.e., the so-called $I_{3322}$
inequality.

Furthermore, we describe  an efficient experimental
method for estimating the Horodecki measure of the CHSH
violation for two polarization qubits in an unknown arbitrary
state. By assuming we have access simultaneously to two
copies of the state, we show how to perform such a measurement
with only six discrete measurement settings for estimating the Horodecki measure.
Of course, with an \emph{a priori} (even partial) knowledge
of a given state or, in particular, of the optimal experimental
settings for its detection, one can further simplify the
proposed method. For example, optimized experimental settings
for the best measurement of the CHSH violation for an \emph{a
priori} known class of two-qubit states were studied in
Ref.~\cite{Bellomo10}.

The paper is organized as follows. In Sec.~\ref{sec:def}  we
review some basic definitions used throughout the paper. In
the following Sec.~\ref{sec:extreme} we provide the boundary
states for a given value of the CHSH violation and analytic
expressions for their entanglement in terms of the degree of CHSH
violation. The extremality conditions for the
negativity and concurrence versus the CHSH violation are
tested in Sec.~\ref{sec:negativity} and
Appendix~\ref{sec:concurrence}, respectively. {In
Sec.~\ref{sec:Bzero} we compare the concurrence and
negativity for states satisfying the CHSH inequality.}
Section~\ref{sec:setup} presents a description of an
experimental proposal for measuring the {maximal}
CHSH violation degree using the same six settings regardless
of the investigated two-qubit state. We conclude in
Sec.~\ref{sec:conclusion}.

\section{ Preliminaries \label{sec:def}}

Throughout this paper we study correlations in two-qubit
systems described by density matrices $\rho$, which can be
expressed in the standard Bloch representation as follows
\begin{equation} \rho  = \frac{1}{4}(I\otimes
I+\vec{x}\cdot\vec{\sigma}\otimes
I+I\otimes\vec{y}\cdot\vec{\sigma}+\!\!\!\sum
\limits_{n,m=1}^{3}T_{nm}\,\sigma _{n}\otimes \sigma _{m}),
\label{rho}
\end{equation}
where the correlation matrix
$T_{ij}=\mathrm{Tr}[\rho(\sigma_{i}\otimes\sigma_{j})]$, and the
vectors $\vec{\sigma}=[\sigma_{1},\sigma_{2},\sigma_{3}]$, and
$\vec{x}$ ($\vec{y}$) with elements
$x_{i}=\mathrm{Tr[}\rho(\sigma_{i}\otimes I)]$
($y_{i}=\mathrm{Tr[}\rho(I\otimes\sigma_{i})]$), are
expressed in terms of the Pauli matrices. As discussed
further in the text, this form of two-qubit density matrix is
very convenient for investigating the CHSH violation.

\subsection{Measure of CHSH violation}

{The CHSH inequality for a two-qubit state {
$\rho\equiv\rho_{AB}$} can be written
as~\cite{Clauser69,Horodecki09}:
\begin{equation}
|\mathrm{Tr}\,(\rho \,{\mathcal B}_{\mathrm{CHSH}})|\leq 2
\label{N08}
\end{equation}
in terms of the CHSH operator
\begin{equation}
{\mathcal B}_{\mathrm{CHSH}}= \vec{a} \cdot \vec{\sigma}
\otimes (\vec{ b}+\vec{b}^{\prime })\cdot \vec{\sigma
}+\vec{a}^{\prime }\cdot \vec{\sigma }\otimes
(\vec{b}-\vec{b}^{\prime })\cdot \vec{\sigma } \label{N09}
\end{equation}
where $\vec{a},\,\vec{a}^{\prime }$ and $\vec{b},
\,\vec{b}^{\prime }$ are unit vectors describing the
measurements (i.e., detector settings) on sides {$A$} (Alice)
and {$B$} (Bob), respectively. As shown by Horodecki
\etal~ \cite{Horodecki95} by optimizing the vectors
$\vec{a},\,\vec{a}^{\prime },\,\vec{b}, \,\vec{b}^{\prime }$,
the maximum possible average value of the Bell operator for
the state $\rho $ is given by {
\begin{equation}
\max_{{\mathcal B}_{\mathrm{CHSH}}}\,\mathrm{|Tr}\,(\rho
\,{\mathcal B}_{ \mathrm{CHSH}})|=2\,\sqrt{\mathcal{M}(\rho )},
\label{N11}
\end{equation}}
where {$\mathcal{M}(\rho)=\max_{j<k}\;\{h_{j}+h_{k}\}\leq 2,$} and
$h_{j}$ $(\,j=1,2,3)$ are the eigenvalues of the matrix
$U=T^{T}\,T$ constructed from the correlation matrix $T$ and
its transpose $T^{T}$. The CHSH inequality is violated if and
only if {$\mathcal{M}(\rho)>1$}~\cite{Horodecki95}. In order to quantify
the violation of the CHSH inequality, one can use {$\mathcal{M}(\rho)$}
or, equivalently,}{
\begin{equation}
B(\rho )\equiv \sqrt{ \max \,[0,\,\mathcal{M}(\rho)-1]}, \label{BIV}
\end{equation}}
which yields $B=0$ if the CHSH inequality is not violated and
$B=1$ for its maximal violation. In  Sec.~\ref{sec:setup} we
will describe how to measure the symmetric  matrix $T^{T}T$
{providing a method for  efficient estimation of $B$ and
$\mathcal{M}$}.

\subsection{Entanglement measures}

In our considerations we apply three popular entanglement
measures: negativity, concurrence and REE.

{``Among all entanglement measures negativity arguably is
the best known and most popular tool to quantify bipartite
quantum correlations''~\cite{Eltschka13}. The negativity for
a bipartite state can be defined
as~\cite{Zyczkowski98,Vidal01}:
\begin{equation}
N({\rho})=\max \Big[0,-2\sum_{j}\mu _{j}\Big],
\label{negativity1}
\end{equation}
where the sum is taken over the negative eigenvalues $\mu
_{j}$ of the partially transposed $\rho$ with respect to one
of the subsystems, as denoted by $\rho^{\Gamma}$. In the case
of two qubits, Eq.~(\ref{negativity1}) simplifies to
\begin{equation}
N({\rho})=\max[0,-2\min{\rm eig}(\rho^{\Gamma})],
\label{negativity2}
\end{equation}
since ${\rho }^{\Gamma}$ has at most one negative eigenvalue
in this case.} The negativity is directly related to the
logarithmic negativity, which has a direct physical meaning
of the entanglement cost under PPT
operations~\cite{Audenaert03,Ishizaka04}. However, for
convenience, we use the negativity instead.

For higher-dimensional systems, the negativity has
another important interpretation as an estimator of entangled
dimensions, i.e., how many degrees of freedom of two
subsystems are entangled~\cite{Eltschka13}. We note that the
dimension of Hilbert spaces can also be tested by the
violations of Bell's inequality~\cite{Brunner08}.

The Wootters concurrence~\cite{Wootters98} is defined as
\begin{equation}
C({\rho})=\max \{0,2\max_j\lambda_j-\sum_j\lambda_j\},
\end{equation}
where $\{\lambda^2 _{j}\} = \mathrm{eig}[{\rho }({\sigma
}_{2}\otimes {\sigma }_{2}){\rho}^{\ast }({ \sigma
}_{2}\otimes {\sigma }_{2})]$. This measure is a monotonic
and convex function of the entanglement of
formation~\cite{Bennett1}. As in the case of the
negativity and logarithmic negativity, it is often more convenient
to operate with the concurrence instead of the entanglement
of formation.

The REE is defined as
\begin{equation}
E_R(\rho)={\rm min}_{\sigma\in {\cal D}}S(\rho
||\sigma)=S(\rho ||\sigma_0),
\end{equation}
where $S(\rho ||\sigma)={\rm Tr}\,( \rho \log_2 \rho
-\rho\log_2 \sigma)$ is the relative entropy to be minimized
over a set ${\cal D}$ of separable states
$\sigma$~\cite{Vedral97a,Vedral98}. The REE is used to
distinguish a density matrix $\rho$ from the closest
separable state (CSS)  $\sigma_0$. For pure states, the REE
reduces to the von Neumann entropy of one of the subsystems.
However, the REE is not a true metric, because it is not
symmetric and does not fulfill the triangle inequality.
An analytical formula for $\sigma_0$ (and thus for the REE) for
a given general two-qubit state $\rho$ is very unlikely to be
found~\cite{Miran08b}. Nevertheless, there is a solution of
the inverse problem~\cite{Miran08a}. Probably, the most
efficient numerical method for calculating the REE was
described in Ref.~\cite{Zinchenko10} and, thus, it is used
here.

\section{Extremal entanglement for a given CHSH violation \label{sec:extreme}}

For each of the three entanglement measures listed in
Sec.~\ref{sec:def}, we can ask about the states that are
extremal, i.e., have the maximal or minimal value of one
entanglement measure for a given fixed value of another
entanglement
measure~\cite{Verstraete01b,Wei03,Miran04a,Miran04b,Miran08b}
or the CHSH violation
measure~\cite{Verstraete02,Horst13}. Similarly,
we state a more specific question about the maximal
entanglement for vanishing of any other fixed degree of the
CHSH violation. In this section we show that for all the
above-mentioned entanglement measures, the states of the
highest (lowest) entanglement for a given violation of the CHSH
inequality are in fact the same class of states denoted as
$\rho_{\max}$ ($\rho_{\min}$).

\subsection{Optimal amplitude damped states}

The amplitude-damped states can be defined by~\cite{Horst13}
\begin{equation}
\rho(\alpha,p)=p|\psi_{\alpha}\rangle\langle\psi_\alpha|
+(1-p)|00\rangle\langle00|, \label{ADS}
\end{equation}
where
$|\psi_{\alpha}\rangle=\sqrt{\alpha}|01\rangle+\sqrt{1-\alpha}|10\rangle$
with $p,\alpha\in[0,1]$. As discussed in detail in
Ref.~\cite{Horst13}, these states can be obtained by
subjecting pure states $|\psi_{\alpha'}\rangle$ to amplitude
damping. {In the special case for $\alpha=1/2$, the state
$\rho(\alpha,p)$ is referred to as the Horodecki state, which
is  a mixture of a Bell state (in our case, the singlet
state) and a separable state orthogonal to it.}

\begin{figure}
\fig{\includegraphics[scale=0.85]{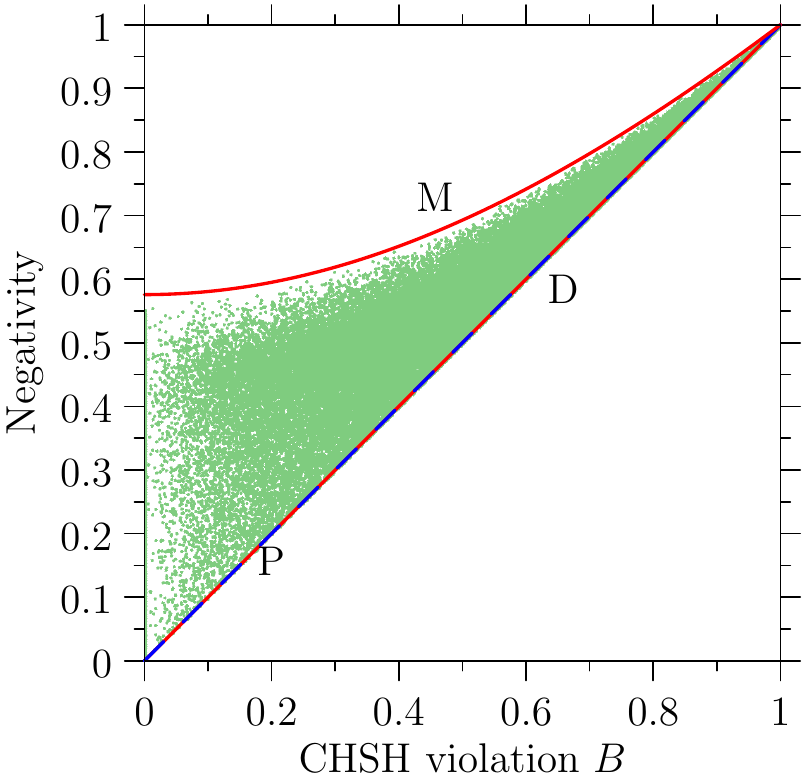}
\includegraphics[scale=0.85]{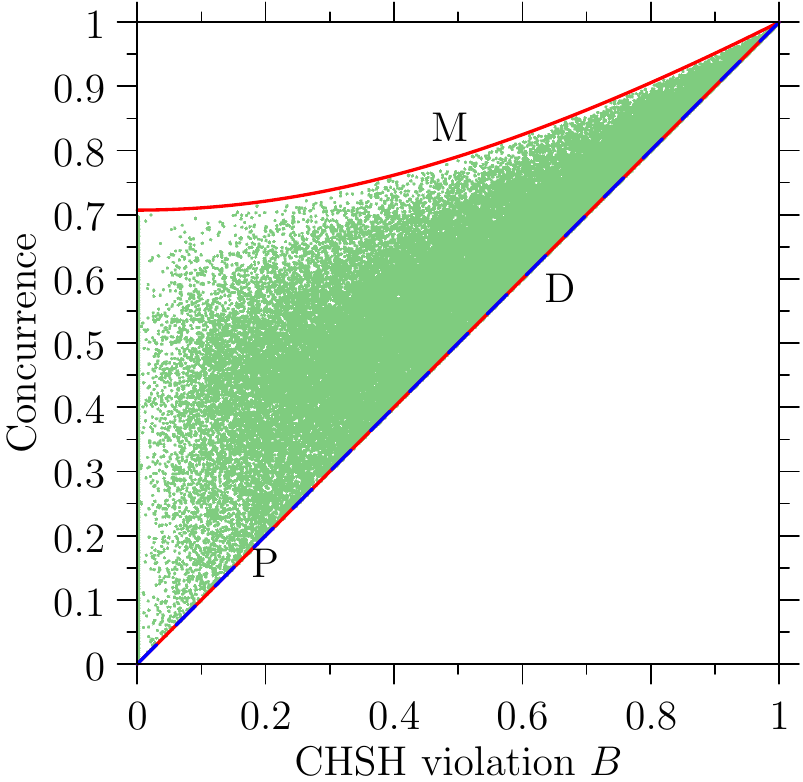}
\includegraphics[scale=0.85]{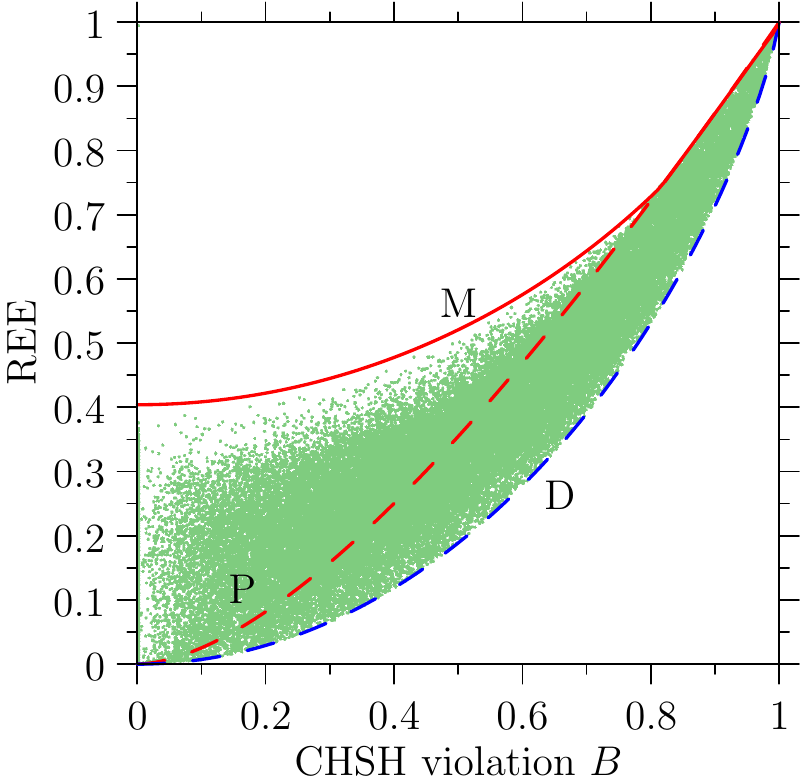}}
\caption{\label{fig:1} (Color online) From the top: the
negativity $N$, concurrence $C$, and relative entropy of
entanglement $E_R$ versus the CHSH violation $B$ for $10^6$
random two-qubit states. The extremal states are marked as P
for pure state, D for Bell-diagonal states $\rho_{\min}$, and
M for $\rho_{\max}$. The maximal values of $N_{\max}(B=0)=
0.57567$, $C_{\max}(B=0)= 0.70711$, and
$E_{R,\max}(B=0)=0.404$ are reached for the M states
$\rho_{\max}$.} \end{figure}

\begin{figure}
\fig{\includegraphics[scale=1]{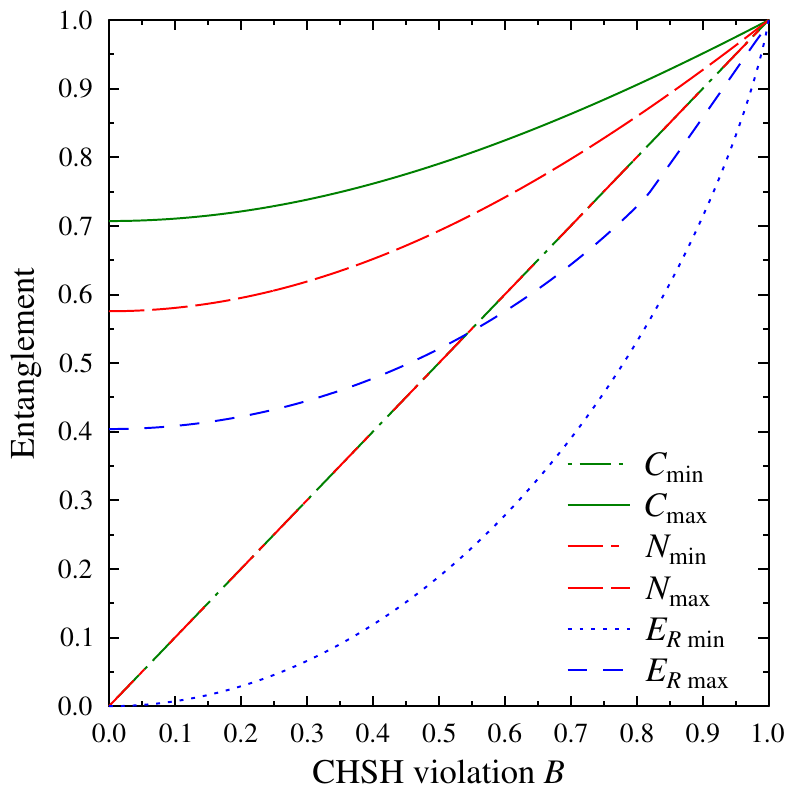}}
\caption{\label{fig:2} (Color online) Comparison of the
extremal values of the relative entropy of entanglement
$E_R$, concurrence $C$, and negativity $N$ as  functions of
the CHSH violation $B$.}
\end{figure}

The amplitude-damped states, which provide the upper bound
for the REE for a given value $B$ of the CHSH violation, are shown in
Figs.~\ref{fig:1} (bottom) and~\ref{fig:2}. Moreover, as found
in Ref.~\cite{Horst13}, the amplitude-damped states
$\rho(\alpha,p)\equiv\rho_{\max}(B_0),$ having the maximal
value of the REE,  $E_{R,\max}$, for a given value $B$, are
the following
\begin{subequations}
\begin{eqnarray} p &=& {\begin{cases}
\frac{1}{4}\left(2 + \sqrt{2 + 2B^2}\right)\qquad\mbox{ if } B <B_0,\\
1\hspace{3.3cm}\mbox{ if } B >B_0, \end{cases} }\\
1&\geq&p\geq
\frac{1}{4}\left(2 + \sqrt{2 + 2B_0^2}\right)\mbox{ if } B = B_0,\\
\alpha &=& \frac{1}{2p}\left(p - \sqrt{5p^2 - 4p -
B^2}\right),
\end{eqnarray}
\label{alpha_opt}
\end{subequations}
where $B_0=0.81686$. For the negativity and concurrence for a
given value of $B$, the states $\rho_{\max}(B_0)$ are also
optimal but for $B_0 = 1$ [so $p=\frac{1}{4}(2 + \sqrt{2 +
2B^2})$ for any $B$] as shown in Sec.~\ref{sec:negativity}
and Appendix~\ref{sec:concurrence}, respectively. Thus,
the maximal values of the entanglement measures for a given
value of {$\xi^2 = B^2 + 1$} are found as
\begin{eqnarray}
N_{\max}(\xi) &=& \frac{\sqrt{2}}{4}\left(\xi +\sqrt{5\xi^2 -
2\sqrt{2}\xi+ 2}\, \right) - \frac{1}{2},
\label{Nmax}\\
C_{\max}(\xi) &=& \frac{\sqrt{2}\xi^2+2\xi}{2\sqrt{\xi^2
+2\sqrt{2}\xi+2}}, \label{Cmax}
\end{eqnarray}
while $E_{R,\max}(B)$ is given in Ref.~\cite{Horst13}.

\subsection{Optimal phase damped states}

The lower bound on the three entanglement measures vs the
CHSH violation $B$ is achieved by the Bell-diagonal states,
{{
\begin{eqnarray}
\rho_{\mathrm{D}}=\sum_{i=1}^{4}\lambda_{i}|\beta_{i}\rangle\langle\beta_{i}|,\label{BD}
\end{eqnarray}}
as labeled by D in Fig.~\ref{fig:1}. Here $|\beta_{i}\rangle$
are the Bell states and the parameters $\lambda_{j}$ are
non-negative and normalized,  $\sum_{i}\lambda_{i}=1$.} The
reason why the Bell-diagonal states provide the lower bound
for $B$ given for a fixed function of the spectral properties
of two-qubit density functions was explained in
Ref.~\cite{Verstraete02}. The Bell-diagonal states can be
produced by two-qubit pure states subjected to phase
damping~\cite{Horst13}. These states, assuming that
$\lambda_3=\lambda_4=0$, can be given in terms of the CHSH
violation degree $B$ as follows:
\begin{equation}
\rho_{\min} = \frac{1}{2}[(1+B)|\beta_1\rangle\langle
\beta_1| + (1-B)|\beta_2\rangle\langle
\beta_2|],\label{rho_min}
\end{equation}
where $|\beta_1\rangle$ and $|\beta_2\rangle$ denote two
orthogonal Bell states. Note that the relation between the
CHSH violation and entanglement for the Bell-diagonal states
is very simple, as given by
\begin{subequations}
\begin{eqnarray}
N_{\min}(B)&=&C_{\min}(B)=B, \\
E_{R{\min}}(B)&=& 1- h[(1+B)/2],
\end{eqnarray}
\end{subequations}
where $h(x) = -x\log_2 x - (1-x)\log_2(1-x)$ is the binary
entropy.

\section{Extremality conditions for negativity for a given
CHSH violation \label{sec:negativity}}

Here we show that the amplitude-damped states given by
Eq.~(\ref{ADS}), for the parameters:
\begin{subequations}
 \begin{eqnarray}
 p &=& \frac{1}{4}\left(2 + \sqrt{2}\xi\right),\\
 \alpha &=& \frac{1}{2}\left(1-\sqrt{1-\frac{4\xi^4}{(\xi^2+\sqrt{2}\xi)^2}}\, \right),
 \end{eqnarray}
\label{ADSparams}
\end{subequations}
where $\xi^2 = 1 + B^2$, are likely to provide the upper
bound of the negativity $N$ for a given value $B$ of the CHSH
violation. For this purpose we apply a generalized method of
Lagrange multipliers and test the KKT
conditions~\cite{Boyd04}.

Thus, let us consider the following Lagrange function
\begin{eqnarray}
\mathcal{L}&=& B(\rho) + l\left[\frac{N}{2}
-\mathrm{Tr}\Big(\rho\psiT\Big)\right]\\ && -
\mathrm{Tr}(X\rho) + \lambda(\mathrm{Tr}\rho-1),
\end{eqnarray}
where $l,\,X$, and $\lambda$ are Lagrange multipliers, and
$\psiT$ is the optimal state for $\rho$ providing
$N(\rho)=-2\mathrm{Tr}[\psiT\rho].$

The Lagrange function is stationary if it remains unchanged
after an arbitrary small deviation of $\rho\to\rho+\Delta$,
where $\Delta$ is defined on the support space of $\rho$.
Thus, our Lagrange function
\begin{eqnarray}
\mathcal{L} &\to & \mathcal{L} +
\mathrm{Tr}\left[\Delta\left(\Bop + l\psiT - X
+\lambda\right)\right],
\end{eqnarray}
should remain constant for small $\Delta$, i.e.,
\begin{subequations}
\begin{eqnarray}
&&\Bop +l\psiT -X+\lambda=0,\label{KKT1} \\ && \hbox{~~~}
X\ge0, \hbox{~~~} \hbox{Tr}(X\rho)=0,
\end{eqnarray}
\end{subequations}
where $\Bop$ is the operator satisfying
$B(\rho)=\mathrm{Tr(}\rho\Bop)$. Let us also note that $X\geq
0$ is required only for the eigenvalues in the support space
of $\rho$.

Moreover, it follows from Eq.~(\ref{KKT1}), after taking the
mean value for $\rho$, that
\begin{equation} \lambda =
l\frac{N}{2}-B(\rho).
\end{equation}
Thus, we can rewrite the KKT conditions as
\begin{eqnarray} \nonumber && X = \Bop -
B(\rho)+l\left(\frac{N(\rho)}{2}+\psiT\right)\geq0,\label{KKT2} \\
&&\mathrm{Tr}(X\rho)=0.
\end{eqnarray}
For the following rank-2 mixed states $\rho =
\lambda_1|e_1\>\<e_1| + \lambda_2 |e_2\>\<e_2|$, which we
conjecture to be extremal on the basis of our numerical
simulation, we can easily derive the following expressions
that can be used with the $l$ multiplier as
\begin{subequations}
\begin{eqnarray}
\langle e_1|\Bop |e_2\rangle &=& - l \langle e_1|\psiT
|e_2\rangle, \label{Condition0} \\ \nonumber \< e_1| \Bop |
e_1\> &=& - l\left(\frac{N(\rho)}{2} + \< e_1|\psiT |
e_1\>\right)\\ &&+ B(\rho).\label{Condition1}
\end{eqnarray}
\end{subequations}
By applying the KKT conditions we can check if a given state
is optimal having its $\Bop$ and $\psiT$.  The CHSH operator
for the amplitude-damped states reads as
\begin{eqnarray}
\label{BIVGH} \Bop = \begin{cases}
\eta_1\left[(1-2p)\sigma_3^{\otimes2} +
2p\sqrt{(1-\alpha)\alpha}\sigma_1^{\otimes2} -1 \right]\\
\qquad
\mbox{     if } 4p^2(1 - \alpha)\alpha - (1 - 2p)^2 <0,\\
\eta_2\left[2p\sqrt{(1-\alpha)\alpha}(\sigma_1^{\otimes2}
+\sigma_2^{\otimes2})-1\right]\\ \qquad\mbox{     otherwise,}
\end{cases}
\end{eqnarray}
where $\eta_1=1/\sqrt{(1-2p)^2+4p^2\alpha(1-\alpha)-1}$ and
$\eta_2=1/\sqrt{8p^2(1-\alpha)\alpha-1}$, whereas
\begin{equation}
|\psi\> = \mathcal{N}\left[(\sqrt{q^2 + 4y^2} - 1)  |00\> +
2y|11\>\right],
\end{equation}
where $y = p\sqrt{\alpha(1-\alpha)}$, $q \equiv 1-p$, and
$\mathcal{N}$ is a normalization constant.

The above results allow us to conclude that the optimal
amplitude-damped states $\rho(\alpha,p)$, which maximize the
negativity $N(\rho)$ for a given $B(\rho)$, are for the
parameters $p$ and $\alpha$, given by Eq.~(\ref{ADSparams}).
These parameters are the same as those resulting in the
maximum REE for a fixed $B$ as given by Eq.~(\ref{alpha_opt})
but with $B_0 = 1$.

Then the negativity for $\rho_{\max}(1)$ can be readily found
as given by Eq.~(\ref{Nmax}), which reaches its maximum
$N_{\max}\approx 0.57567$ for $B=0$. This result is confirmed
by our Monte Carlo simulation shown in Fig.~\ref{fig:1} (top).

Similar reasoning confirms that the minimal negativity is
reached by the $\rho_{\min}$ states, given by
Eq.~(\ref{rho_min}).

\section{Concurrence vs negativity if CHSH inequality is satisfied \label{sec:Bzero}}

\begin{figure}
\fig{\includegraphics[scale=1]{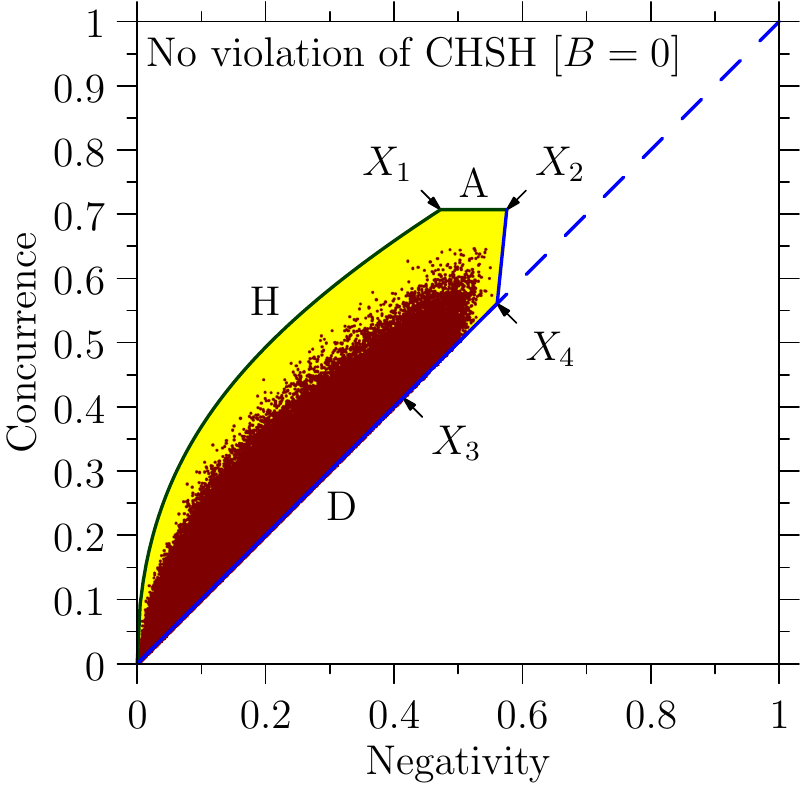}}
\caption{\label{fig:3} {(Color online) Concurrence $C$ vs
negativity $N$ for two-qubit states satisfying the CHSH
inequality ($B=0$). All these states cover the region marked
in bright yellow. States generated by our Monte Carlo simulations
correspond to dark brown dots. The extremal states are marked as D
for the Bell-diagonal states, H for the Horodecki states, and
A for the amplitude-damped states given by Eq.~(\ref{ADS})
with the proper choice of parameters $p,\alpha$. The points
$X_k=(N_k,C_k)$ and other details are specified in the
text.}}
\end{figure}

One can conjecture that there is a direct relation between
the concurrence and negativity for a fixed CHSH violation for
the simple case of general two-qubit mixed states, which
would then make the results of the former sections somewhat
trivial. Here we show that there is no such relation for
general two-qubit mixed states for a fixed $B=0$.

It is worth noting that even if a given entangled state does
not violate any Bell type inequality, but still can be used
for quantum teleportation, as shown by
Popescu~\cite{Popescu94} on the example of Werner's entangled
state given by Eq.~(\ref{eq:1}) for $p\in (1/3,1/\sqrt{2}]$.
Popescu concluded that: ``The nonlocality responsible for
violations of Bell's inequalities is not equivalent to that
used in teleportation, although they probably are two aspects
of the same physical property''~\cite{Popescu94}. However, it
has recently turned out that these two aspects are in fact
quite closely connected. Indeed, all entangled states useful
for teleportation lead to deterministic violation of Bell's
inequality (so, they are nonlocal resources) as shown by
Cavalcanti \etal~ \cite{Cavalcanti13} with the help of the
phenomenon of superactivation of quantum nonlocality.
Moreover, as demonstrated in Ref.~\cite{Masanes06}, all
bipartite entangled states are useful as a nonclassical
resource for quantum information processing.

Figure~\ref{fig:3} shows the area covered by two-qubit states
$\rho$ satisfying the CHSH inequality [i.e, $B(\rho)=0$] for
the concurrence $C(\rho)$ plotted vs the negativity
$N(\rho)$. The marked points $X_k=(N_k,C_k)$, with
$k=1,...,4$, correspond to the following negativities:
\begin{eqnarray}
 N_1&=&\frac{1}{\sqrt{2}}+\sqrt{2-\sqrt{2}}-1\approx 0.4725, \nonumber\\
 N_2&=&\frac14 \Big(\sqrt{2}+\sqrt{14-4\sqrt{2}}-2\Big)\approx 0.5757,
\label{eq:N12}
\end{eqnarray}
$N_3=\sqrt{2}-1\approx0.4142,$ and $N_4=\frac14
(3\sqrt{2}-2)\approx 0.5607$, as well as  to the
concurrences: $C_1=C_2=1/\sqrt{2}\approx 0.7071$, and
$C_k=N_k$ for $k=3,4$. The upper bound of this area for
$N\in[0,N_1]$ is given by
\begin{eqnarray}
C(N)=\sqrt{2N(N+1)}-N, \label{UpperBound1}
\end{eqnarray}
or, equivalently, by $N(C)=\sqrt{(1-C)^2+C^2}-(1-C)$.  This
bound can be reached by the Horodecki states ({labeled by H}),
given by Eq.~(\ref{ADS}) for $\alpha=1/2$ and $p=C(\rho)$. We
note that the upper bound of the concurrence vs negativity
without specifying the CHSH violation is also given by
Eq.~(\ref{UpperBound1}) but for the whole range $0\le N \le
1$~\cite{Verstraete01b,Miran04a}. The upper bound for
$N\in[N_1,N_2]$ is likely to be $C(N)=1/\sqrt{2}$, which can be
reached by the amplitude-damped states ({labeled by A}) given
by Eq.~(\ref{ADS}) for $p=1/[2\sqrt{2\alpha(1-\alpha)}]$ and
$\alpha=[\alpha_-,\alpha_+]$, with
$\alpha_{\pm}=1/2\pm\sqrt{8\sqrt{2}-11}$. The lower bound for
$N\in[0,N_4]$ is simply given by $C=N$ and can be reached by
the rank-4 Bell-diagonal states, given by Eq.~(\ref{BD}) for
$\lambda_1=\lambda_2=\lambda_3=(1-N)/6$ and, thus,
$\lambda_4=\max_n\lambda_n=(1+N)/2$. The lower bound for
$N\in[0,N_3]$ can also be reached by the rank-3 Bell-diagonal
states with $\lambda_1=0$, $\lambda_2=\lambda_3=(1-N)/4$, and
$\lambda_4$ being the same as in the previous case. The rank-2
Bell-diagonal states satisfying $B=0$ correspond only to the
point $C=N=0$. It is also worth noting that the lower bound
of $C$ vs $N$ for arbitrary $B$ is also simply given by $C=N$
but for any $0\le N\le 1$~\cite{Verstraete01b,Miran04a}. We
could not find analytical examples of states corresponding to
the lower bound for $N\in[N_4,N_2]$. Note that it is a very
narrow region, as equal to $N_2-N_4=0.015$, so it is even
difficult to numerically simulate states satisfying both
$N\in(N_4,N_2)$ and $B=0$. Of course, one can analyze, e.g.,
the mixture $\rho_q=q\rho_{X_2}+(1-q)\rho_{X_4}$,
where $0\le q\le 1$, of the discussed Bell-diagonal state at point
$X_4$ and the amplitude-damped state at point $X_2$. Then one
can observe that only for $q\ge 0.9893...$ is the negativity
$N(\rho_q)\ge N_4$, which corresponds to the
concurrence $C(\rho_q)\ge 0.6706
> C_4$. This explicitly shows that the state $\rho_q$ can have the
negativity $N\in(N_4,N_2)$ but it is not the lower bound of
$C$ vs $N$ for $q>0$.

A closer analysis of Fig.~\ref{fig:3} also shows that there
are infinitely many pairs of two-qubit states (say, $\rho_1$
and $\rho_2$) \emph{violating} the following intuitive
conditions for ordering states with the concurrence and
negativity:
\begin{eqnarray}
 N(\rho_1) = N(\rho_2) \Leftrightarrow C(\rho_1) = C(\rho_2) \nonumber \\
 N(\rho_1) > N(\rho_2) \Leftrightarrow C(\rho_1) > C(\rho_2) \nonumber \\
\end{eqnarray}
for a fixed CHSH violation $B(\rho_1) = B(\rho_2)$. Of
course, there are also infinitely many other states
satisfying these conditions. In particular, by analyzing
Fig.~\ref{fig:3}, one can find analytical nontrivial examples
of states which  satisfy the CHSH inequality and are ordered
differently by these entanglement measures, e.g.,
\begin{eqnarray*}
N(\rho_{X_1}) < N(\rho_{X_4}) \quad &\mbox{and} \quad C(\rho_{X_1}) >C(\rho_{X_4}), \label{order2}\\
N(\rho_{X_1}) =N(\rho_{X_5}) \quad &\mbox{and} \quad C(\rho_{X_1}) > C(\rho_{X_5}), \label{order3}\\
N(\rho_{X_1}) < N(\rho_{X_2}) \quad &\mbox{and} \quad C(\rho_{X_1})
= C(\rho_{X_2}), \label{order4}
\end{eqnarray*}
where $\rho_{X_5}$ is, e.g., the Bell-diagonal state with
$\lambda_1=\lambda_2=\lambda_3=(1-N_1)/6$,
$\lambda_4=\max_n\lambda_n=(1+N_1)/2$, and $N_1$ the same as
for $\rho_{X_1}$. Of course, one can also identify pairs of
states, e.g., $\rho_{X_2}$ and $\rho_{X_4}$, which are ordered
in the same way, e.g., $N(\rho_{X_2})
> N(\rho_{X_4})$ and $C(\rho_{X_2}) >C(\rho_{X_4})$.

This relativity of ordering states by different entanglement
measures is a well-known
phenomenon~\cite{Eisert99,Zyczkowski99,Virmani00,Wei03,Miran04a,Miran04b,Miran04c,Miran04d},
which clearly shows the lack of simple relations of, e.g.,
the negativity and concurrence (when the degree of the CHSH
violation is irrelevant) for  two-qubit states. Here we
showed the  relativity of ordering states by  the negativity
and concurrence for a fixed value of the CHSH violation.

\section{Proposal of efficient
measurement of the correlation matrix $T^TT$ \label{sec:setup}}

\begin{figure}
\fig{\includegraphics[width=8cm]{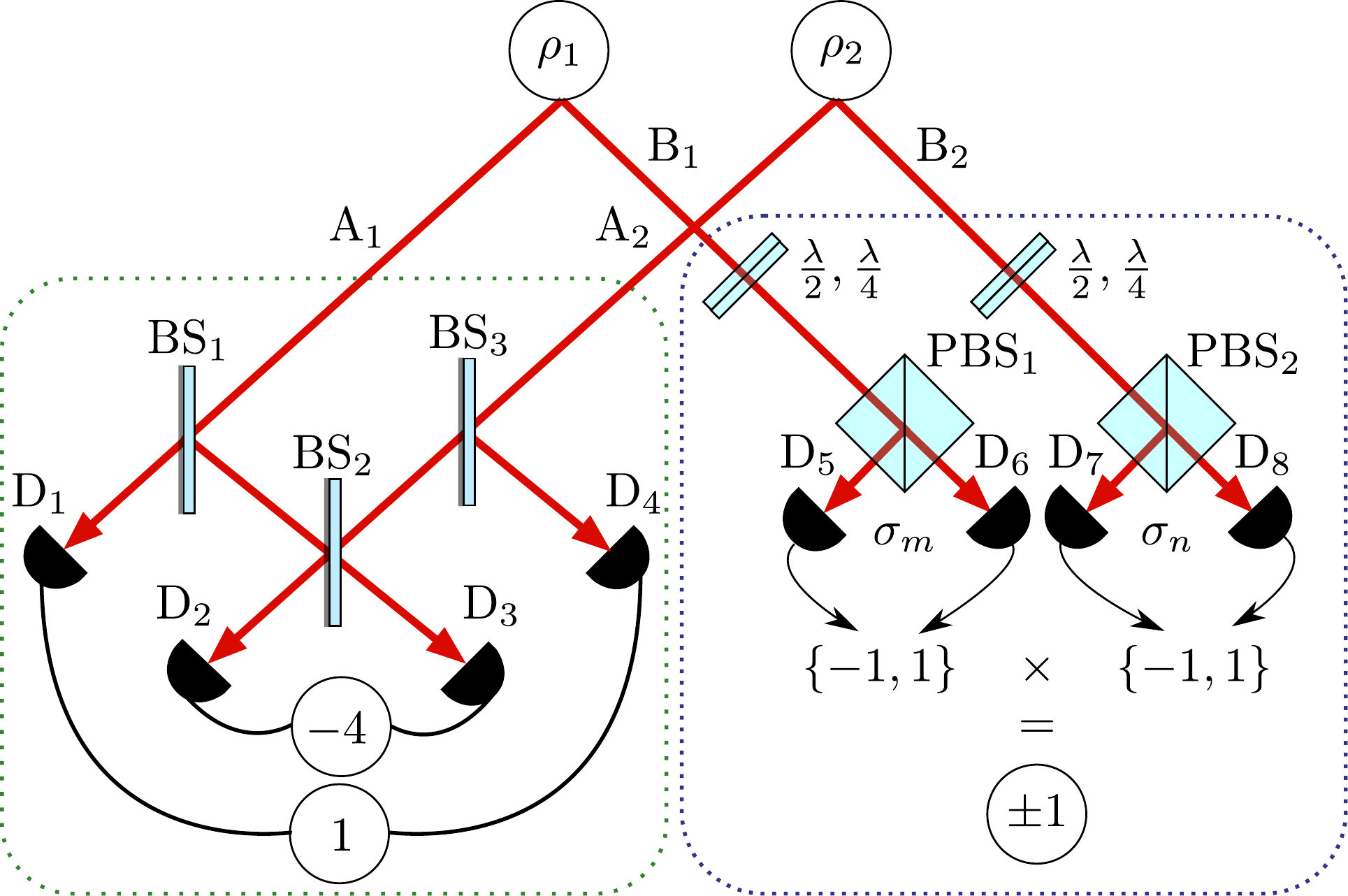}}
\caption{\label{fig:setup} (Color online)  Setup implementing
the measurement of $(T^TT)_{m,n}$ using two sources (or a
single photon source with routing and delaying every second
pair of photons) of a two-qubit state ($\rho_1$ and
$\rho_2$). The basic building blocks are as follows: beam
splitters (BSs), polarizing beam splitters (PBSs),
quarter-wave plate ($\lambda/4$) and half-wave plate
($\lambda/2$), and standard detectors. The values of
$m,n=x,y,z$ are set by rotating the polarization  by means of
the wave plates, i.e., one $\lambda/4$ and one $\lambda/2$
plate, where $\lambda$ is the wavelength. Circled $-4$ (and $\pm
1$) means that this value is assigned if the corresponding
detectors (D) click. Since the investigated function of the
correlation matrix $T$ is symmetric we need only to measure
it in six configurations, e.g, $(m,n) = (x,x), (x,y),
(x,z),(y,y), (y,z), (z,z)$. Due to the probabilistic nature  of
the path taken by photons after the BS interaction, the setup
gives a conclusive result in  half of the cases if $\rho_1$ and
$\rho_2$ are supplied at the input. }
\end{figure}

Knowing the lower and upper bounds of the three entanglement
measures for a given CHSH violation $B$, we are now able to
deduce the range of entanglement of any two-qubit state for a
fixed $B$. The expression for the CHSH violation $B$ depends
solely on the eigenvalues of the symmetric real matrix
$T^TT$. Here we present an efficient method for measuring
this correlation matrix $T^TT$.

We can express the elements of this matrix using two copies
$\rho_1$ and $\rho_2$ of the two-qubit state $\rho$ as
\begin{equation}
(T^TT)_{m,n}=\mathrm{Tr}\left[(\rho_{A_1B_1}\otimes
    \rho_{A_2B_2})U_{A_1A_2}\otimes(\sigma_m\otimes\sigma_n)_{B_1B_2}\right] ,\label{TTT}
\end{equation}
where $\rho_{A_1B_1}\equiv\rho_1$ and {
$\rho_{A_2B_2}\equiv\rho_2$} for the subsystems $A$ and $B$,
whereas the operator $U_{A_1A_2} = (-4|
\Psi^-\rangle\langle\Psi^-| + I)_{A_1A_2}$ is given in terms
of the singlet projection $|\Psi^-\rangle\langle\Psi^-|$ onto
the corresponding subsystems and the two-qubit identity operation $I$
(for a derivation see Appendix~\ref{sec:TTT}). Since the
$3\times 3$ matrix $T^TT$ is symmetric, $(T^TT)_{m,n} =
(T^TT)_{n,m}$, so it is completely defined by six real
numbers, which can be directly measured  for, e.g.,
single-photon polarization qubits. We choose, e.g.,
$|0\rangle$ ($|1\rangle$) to represent a horizontally
(vertically) polarized photon. For such qubits, $T^TT$ can be
measured by the setup shown in Fig.~\ref{fig:setup}. The
left-hand-side module of this setup, which consists of three
50:50 asymmetric beam splitters (BSs), performs the
measurement of the $U_{A1A2}$ operator. The operation of this
module was described in detail (considering imperfections
including finite detection efficiency) in
Ref.~\cite{Bartkiewicz13}. The possible outcomes for a single
measurement instance $a_k$ are 
$a_k \in \lbrace -4,-1,0,1,4 \rbrace$, which are
the products of the outcomes $-4,0,1$ corresponding to a
particular coincidence detection in the module $1$, for a
coincidence detection in the detectors {D$_1$} and {D$_4$}; $-4$,
for a coincidence detection in the detectors {D$_2$} and
{D$_3$}; and $0$, if neither of the two coincidences has been
detected. Moreover, the right-hand-side module of the setup
in Fig.~\ref{fig:setup} measures the product
$\sigma_m\otimes\sigma_n$.  The outcomes $-1,1$ of this
module occur for measuring the product of the Pauli matrices.
The useful values of $a_n\neq0$ appear for one-half of the
cases when the states $\rho_1$ and $\rho_2$ are delivered and
assuming perfect detectors. For realistic components (see the
analysis in Ref.~\cite{Bartkiewicz13}), this setup would
provide us with a good estimation of $T^TT$ in a time period
corresponding to switching between the six settings of
$\sigma_m\otimes\sigma_n$ instead of nine settings required
for the full tomography of the $T$ matrix. The 
expected values obtained read
\begin{equation}
\langle\sigma_m\otimes\sigma_n\rangle  =
\frac{1}{K_0}\sum_{k=1}^K a_k,
\end{equation}
where $K$ is the number of measurements, $K_0 = \sum_{k=1}^{K}\delta_{|a_k|,1}$ and
$\delta_{|a_k|,1}$ is the Kronecker $\delta$. Note that the
depicted measurement method is not limited to measuring $B$
for any two-qubit state. It measures $T^TT$, which contains
more information than the sum of the two largest eigenvalues
used for calculating $B$.

\subsection{Proposal for experimental optimization \label{sec:experiment}}

Here we discuss  an optimization of the setup to make it
experimentally more feasible. Our implementation of the
left-hand-side module, depicted conceptually  in
Fig.~\ref{fig:setup}, requires three balanced beam splitters
and four detectors. From the experimentalist point of view,
the larger the number of components, the larger the
measurement error. For example, the splitting ratio of beam
splitters is particularly sensitive to mount alignment and
manufacturing precision. Furthermore all the detectors have
to be calibrated to the same relative detection efficiency.

\begin{figure}
\fig{\includegraphics[width=3.0cm]{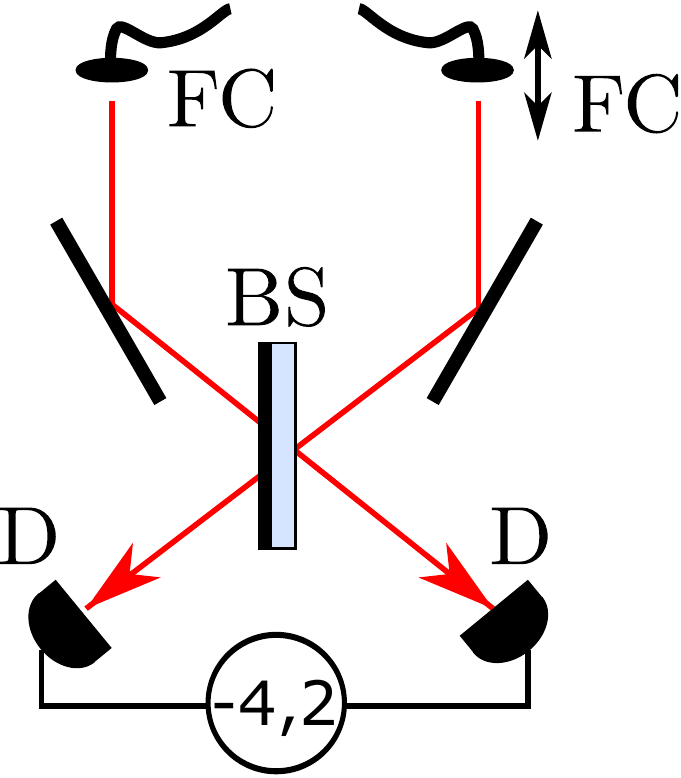}}
\caption{\label{fig:experimental} (Color online)
Experiment-friendly setup replacing the left-hand side
measurement module in Fig.~\ref{fig:setup}. BS, balanced
beam splitter; FC, fiber coupler; D, detector. Motorized
translation (marked by double arrow) is used to tune the
temporal delay between the photons in order to switch between
measurement regimes as explained in the text.}
\end{figure}

To reduce the number of required optical components, we
propose a modified measurement setup depicted in
Fig.~\ref{fig:experimental}. As the conceptual setup in
Fig.~\ref{fig:setup} shows, there are two distinct
measurement regimes in the module. The first regime is
implemented by two-photon overlap on a balanced beam splitter
projecting the state onto the singlet state. The
corresponding coincidence rate is then multiplied by the
factor of $-4$. The second regime is just a plain coincidence
count (detectors D$_1$ and D$_4$). In the modified setup, we
implement both these regimes using a single beam splitter. To
switch between the regimes, we suggest a delay line to tune
the temporal overlap between the interacting photons. The
first measurement regime is obtained by setting the delay
between the photons to zero, while the second regime is
obtained when the delay is sufficiently larger than a
single-photon coherence length. In the second regime, the two
photons impinge on the beam splitter independently and they
exit by different output ports in half of the cases only. For
this reason, this number of coincidences has to be multiplied
by 2 to implement the conceptual setup.

The benefits of the optimized setup are at least threefold:
(i) The number of beam splitters is reduced by a factor of
3. (ii) Only one pair of detectors is used, so there
is no need for the calibration within the module. The only
calibration to be performed is the mutual calibration of the
efficiencies of the detector pairs across the left- and right-hand-side
modules of the setup. (iii) Another minor benefit
of the modified version of the setup is that it can be
constructed using a standardized two-photon-state
characterization device~\cite{Halenkova12} routinely used in
other experiments. Note that since there is need for
singlet-state projection even in the original setup, such
a delay line would be needed in order to stabilize the setup
anyway. Therefore it does not impose any additional
experimental requirements.

\section{Conclusions \label{sec:conclusion}}

We have analyzed, as summarized in Fig.~\ref{fig:2}, the
relation between the Horodecki measure of the CHSH inequality
violation (or single-copy nonlocality) and three common
entanglement measures: the negativity, concurrence, and
relative entropy of entanglement. We discovered optimal
states that provide the upper bound on the entanglement
measures for a given CHSH violation. We provided both
numerical and analytic evidence by testing the KKT
extremality conditions within a generalized Lagrange
multiplier method in the case of the negativity for a given
CHSH violation.  We also checked that the states found
satisfy the Verstraete-Wolf conditions~\cite{Verstraete02}
for the extremal concurrence for a given CHSH violation.
Remarkably, the states belong to the same class 
of states for all the investigated
measures of entanglement, including the REE. We showed that
the states providing  the upper and lower bounds on the
entanglement measures for a given value of the CHSH violation
can be simply obtained by the amplitude and phase damping of
pure states, respectively. {We also found extremal states
for the concurrence versus negativity for a fixed value of the
CHSH violation (i.e., $B=0$).}

Moreover, we described a method to efficiently measure
the correlation matrix $T^TT,$ and, thus, to estimate the
Horodecki degree of the CHSH violation. This method together
with the found bounds on the entanglement measures discussed
provides an easy and practical way of estimating entanglement
for arbitrary two-qubit states with a fixed degree of 
CHSH violation.

It is worth comparing our method with the standard methods,
in which the violation of the CHSH inequality can be tested
using four correlation measurements. Hence, one could ask
about the advantage in estimating (in a non
device-independent way) a correlation matrix using six
correlation measurements. One might think that we
use more measurements to achieve less. However, this
four-measurement approach refers  just to \emph{testing} the
CHSH violation for a given state and for given positions of
analyzers. In contrast, our work is about \emph{quantifying}
the CHSH violation for a given state by \emph{optimizing}
over all possible positions of analyzers to have the greatest
degree of the CHSH violation. This approach requires more
measurements than in the case of ordinary unoptimized
measurements of the CHSH violation. Namely, our approach is
based on the Horodecki \emph{measure} of the CHSH violation
corresponding to finding eigenvalues of a real symmetric $3\times3$
matrix with six independent unknown parameters for a given
two-qubit state. Thus, one can conjecture that the minimum
number of optical measurements is six, at least, if  two
copies of the state are simultaneously~available
\cite{Pawel_Horodecki}. If only one copy is available at a
given moment, then the required number of measurements is
even higher (arguably, equal to nine \cite{Pawel_Horodecki}).

Both upper and lower bounds are operationally important
especially in relation to secure quantum communication (for a
related study of secure quantum teleportation see, e.g.,
Ref.~\cite{Ozdemir07}). For example, let us assume that the
degree of CHSH violation (including the case of no
violation) of a given state $\rho$ is known. Then by applying
our negativity bounds, we can calculate the bounds on the
PPT-entanglement cost, which is the asymptotic number of
maximally entangled states that are required to create the
state $\rho$ under operations preserving the positivity of the
partial transpose. Analogously, by applying the
Verstraete-Wolf concurrence bounds, one can calculate the
bounds on the entanglement of formation.

It is worth noting that by measuring the correlation matrix
$T$ of an arbitrary unknown two-qubit state, we can find
directly the optimal measurement settings by applying the
Horodecki theorem~\cite{Horodecki95}. In contrast, by
measuring $T^T T$ of a general state, we can determine the
value of the CHSH violation optimized over all possible
measurement settings without knowing these optimal settings
explicitly. Only for a limited class of states, including
those with symmetric $T$,  the optimal settings can be
determined completely from the $T^T T$ matrix.

Finally, we mention one possible application of our results.
Recently, the CHSH inequality has been proved extremely
useful for verifying the quantumness of a black box device (say,
a claimed quantum computer) programmed to win the so-called
CHSH game~\cite{Reichardt13}. Thus, with the help of our
results, by looking at the results of the CHSH game, we are
able to estimate how much entanglement was used by the tested
black box.

\begin{acknowledgments} {We thank  Pawe\l{} Horodecki, Yeong-Cherng Liang, and Satoshi Ishizaka for
discussions.} This work was supported by the Polish National
Science Center under Grants No. DEC-2011/03/B/ST2/01903 and
No. DEC-2011/02/A/ST2/00305. K.~B. gratefully acknowledges
support by the Operational Program Research and Development
for Innovations - European Regional Development Fund
(Project No. CZ.1.05/2.1.00/03.0058) and the Operational Program
Education for Competitiveness -- European Social Fund (Project No.
CZ.1.07/2.3.00/20.0017 and No. CZ.1.07/2.3.00/30.0041) of the Ministry of Education, Youth and Sports of the Czech Republic.
K.~L. acknowledges the support by the Czech Science Foundation (Project No. 13-31000P).
\end{acknowledgments}

\appendix

\section{States with extremal concurrence for a given CHSH violation
\label{sec:concurrence}}

The conditions satisfied by the extremal amount of the CHSH
violation for a fixed value of the concurrence were given  by
Verstraete and Wolf in Ref.~\cite{Verstraete02}. Note that,
for quantifying the CHSH violation, the authors of
Ref.~\cite{Verstraete02} used the  parameter $\beta
= 2\sqrt{B^2 + 1}$. So, the CHSH inequality is satisfied for
$\beta\le 2$. Nevertheless, their results are valid also in
our case since $B$ is uniquely determined by $\beta$. In
order to solve the optimization problem, the method of 
Lorentz transformations on the extended correlation matrix $T$
was used in Ref.~\cite{Verstraete02} to generate states of
constant concurrence. It was found that pure and
Bell-diagonal states have the maximal concurrence for a given
value $B$ of CHSH violation, while the lowest concurrence
for a given $B$ is achieved by, e.g., a mixture of a Bell
state and a separable state orthogonal to it (the so-called
Horodecki state). A summary of these results is shown in
Fig.~\ref{fig:1}(middle).

The optimality of the states $\rho_{\max}$, given by
Eqs.~(\ref{ADS}) and~(\ref{ADSparams}), for the whole range
of $B$ can be demonstrated using the optimality conditions
given in Ref.~\cite{Verstraete02}. This is straightforward
since the matrix $R_{m,n}=\<\sigma_m\otimes\sigma_n\>$ (for
$m,n=0,1,2,3$, where $\sigma_0$ is the identity), which was used for testing the optimality
conditions in Ref.~\cite{Verstraete02},  has the same
structure as $\rho_{\max}$. The relevant parameters as
defined in Ref.~\cite{Verstraete02} read
\begin{subequations}
\begin{eqnarray}
a^{(\pm)} &=& -\frac{\sqrt{2} (\xi^{2} -2) }{4
(\xi+\sqrt{2})}\pm \frac{ \sqrt{2}}{4}\sqrt{\xi^2
+2\sqrt{2}\xi+2},\\  x &=&y= \frac{\sqrt{2}}{2}\xi, \qquad z
= -\frac{\sqrt{2}\xi+\xi^2}{\sqrt{2}\xi+2},
\end{eqnarray}
\end{subequations}
where $\xi^2 = B^2 + 1$. These parameters satisfy the
optimality conditions (in fact, they saturate the last two):
\begin{subequations}
\begin{eqnarray}
-1 &\leq & z \leq 1,\\ (1+z)^2 - (a^{(+)}+a^{(-)})^2 &\geq &
(x-y)^2,\\ (1-z)^2 - (a^{(+)}-a^{(-)})^2 &\geq & (x+y)^2.
\end{eqnarray}
\end{subequations}
Moreover, the concurrence for  amplitude-damped states is
\begin{equation} C(\alpha,p) = 2p\sqrt{\alpha(1-\alpha)},
\end{equation} so, in the case of extremal states, it can be
expressed by Eq.~(\ref{Cmax}). Thus, the states
$\rho_{\max}$ belong to the class of states having the
highest concurrence for a given degree $B$ of the CHSH
violation reaching the maximum $C_{\max}(\xi=1)\equiv
C_{\max}(B=0) = 1/\sqrt{2}$ as shown in
Fig.~\ref{fig:1}(middle).

\section{Two-copy formula for correlation matrix $T^TT$ \label{sec:TTT}}

Here we derive a two-copy formula for the correlation matrix
$T^TT$, given by Eq.~(\ref{TTT}), which is useful for our
experimental proposal.

In the following, we use the Einstein summation convention.
Let us start by recalling that we can express $T_{mn}$ as
an expectation value of the Pauli matrices, i.e.,
\begin{equation}
T_{mn} = \mathrm{Tr}[(\sigma_m\otimes\sigma_n)\rho],
\end{equation}
hence
\begin{eqnarray}
(T^TT)_{mn}&=&T_{km}T_{kn} =\mathrm{Tr}\nonumber
[(\sigma_k\otimes\sigma_m\otimes\sigma_k\otimes\sigma_n)
(\rho\otimes\rho)]\\\nonumber &=& \mathrm{Tr}
[(\sigma_k\otimes\sigma_k)\otimes(\sigma_m\otimes\sigma_n)
(\rho\otimes\rho)']\\
&=& \mathrm{Tr} \lbrace [U_{A_1A_2}
\otimes(\sigma_m\otimes\sigma_n)_{B_1B_2}]'
(\rho\otimes\rho)\rbrace , \label{B2}
\end{eqnarray}
where
$(\rho\otimes\rho)'=S_{A_2B_1}(\rho\otimes\rho)S_{A_2B_1}$,
$U = \sigma_k\sigma_k=I -4|\Psi^-\rangle\langle\Psi^-|$, and
$|\Psi^-\rangle$ denotes the singlet state. The unitary
transformation $S_{A_2B_1}=I\otimes S\otimes I$ swaps the
modes $A_2$ and $B_1$, which can be given in terms of the
swap operator
\begin{equation}
S =\left( \begin{array}{cccc}
1 & 0 & 0 & 0\\
0 & 0 & 1 & 0\\
0 & 1 & 0 & 0\\
0 & 0 & 0 & 1
\end{array}\right).
\end{equation}
Equation~(\ref{B2}) finally results in Eq.~(\ref{TTT}).


\end{document}